\documentclass[final,5p,times,twocolumn]{elsarticle}
\newcommand{\pf}[2]{\frac{\partial ,1}{\partial ,2}}
\newcommand{\pfs}[2]{\frac{\partial^2 ,1}{\partial ,2 ^2}}
\newcommand{\pft}[2]{\frac{\partial^3 ,1}{\partial ,2 ^3}}
\newcommand{\pderiv}[2]{\frac{\partial ,1}{\partial ,2}}

\usepackage{multirow}
\usepackage{tabu}
\usepackage{CJKutf8}
\usepackage[T1]{fontenc}
\usepackage[utf8]{inputenc}
\usepackage{graphics, amsmath}
\usepackage{graphicx,color}
\usepackage{epsfig}
\usepackage{amsmath}
\usepackage{amssymb}
\usepackage[mathlines]{lineno}
\usepackage{setspace}
\usepackage{hyperref}
\usepackage{cleveref}
\usepackage{underscore}
\usepackage[dvipsnames]{xcolor}
\usepackage{multirow, makecell}
\usepackage{colortbl}
\usepackage{dashrule}
\usepackage{ehhline}
\usepackage{csquotes}
\usepackage{natbib}
\biboptions{sort&compress}
\usepackage{array}
\usepackage{booktabs}
\usepackage{caption}
\makeatletter
\usepackage{colortbl}
\usepackage{hhline}

\journal{International Communication in Heat and Mass Transfer}

\begin{document}

\begin{frontmatter}
	\title{The Non-Local Dual Phase Lag Model of Heat Conduction in a Silicon Metal-Oxide-Semiconductor Field-Effect Transistor}
	\author[1]{Sharif A. Sulaiman}
\author[1]{Zahra Shomali\corref{cor1}}
\address[1]{Department of Physics, Basic Sciences Faculty, Tarbiat Modares University, Tehran, Iran}
	\cortext[cor1]{Corresponding author, Tel.: +98(21) 82884785.}
\ead{shomali@modares.ac.ir}

\begin{abstract}
	
As the transistors and consequently the chips are getting smaller, the accurate investigation of heat transport at micro/nanoscale, becomes an important issue of concern. This is due to an increase in the energy consumption and the leakage currents as a result of the miniaturization which requires taking care of the thermal behavior to make sure that the device is working in the threshold temperature regime. The current work deals with a two-dimensional framework, incorporating the non-locality in space, for more accurate investigation of the micro/nanoscale heat transport using the lower computational cost phenomenological macroscopical Dual Phase Lag (DPL) method. The non-dimensional non-locality parameter $\gamma$, which indicates the strength of the nonlocality, is embedded through the modified DPL model named as non-local DPL (NDPL). In the present manuscript the two-dimensional silicon transistor, has been investigated. It is obtained that the $\gamma$ parameter in x and y direction has the same value and like its behavior at one-dimension, is linearly dependent on the Knudsen number, Kn, being 1.5 for Kn=10 and 0.015 for Kn=0.1. Also, the phase lagging ratio, B, is found to be unique value of 0.08 for all specified times and Knudsen numbers. It should be mentioned that although the effect of non-locality is more pronounced for smaller systems with higher Knudsen numbers in which the non-Fourier behavior is more evident, but contemplating the non-locality parameter in systems with lower Knudsen number, makes the results more precise. When the size of the layer in x-dimension, is two times the y-dimension, it is found that the temperature contour plot has an asymmetric behavior which is more pronounced at the x-direction due to the more accumulation of the non-locality effect. In fact, the silicon is thermally isotropic and so the non-symmetric behavior is attributed to the geometry of the device and the existence of non-locality parameters at x and y directions. In brief, it is affirmed that taking into account the $\gamma$ parameter is noteworthy for accurately predicting the thermal behavior in micro/nano scale systems using the classical macroscopical methods.

\end{abstract}


\end{frontmatter}

\section{Introduction}
\label{introduction}
Micro/nano-scale heat conduction in nano-materials, especially the semiconductors, has been the matter of interest in recent years \cite{Samian2013,Samian2014,Moghadam2014,Shomali2019}. This concern on one hand, relates to the practical application in nano-electrics, and on the other hand, brings out valuable understanding of the fundamental principles \cite{Shomali2018,Bahadori2024}. In fact, the famous Fourier's law, $\vec{q}(\vec{r},t)=-k\vec{\nabla}T(\vec{r},t)$, which is known as the accurate model to simulate heat conduction in macroscopics structures, fails when length/time scales are, respectively, comparable to the phonon mean free path and the relaxation times. In fact, there exist atomistic and also phenomenological macroscopic methods for heat transport study in micro/nano-scale devices. The atomistic models are more complicated, computationally expensive and simultaneously very accurate. To provide an estimate of the computational cost, one can consider solving the phonon Boltzmann transport equation (PBE), which is an atomistic method, to investigate the thermal transport in nanodevices. Using this approach, we have simulated the transient thermal behavior of a 3-D silicon of 10$\times$10$\times$1000 nm$^3$, for 1 ns. It is obtained that using the 8 cores with the specifications of 2:8 GHz and 32 GB RAM, one needs 2 days, 13 h and 29 min for completing the simulation. This observation shows how computationally time-consuming are atomistic methods.
 	 On the other hand, the macroscopic heat transfer models in support of less computational cost and not very precise but acceptable results are introduced. Actually, several non-Fourier heat conduction constitutive laws have been proposed. For instance, one can mention the Cattaneo-Vernotte (CV) \cite{Cattaneo1958,Vernotte1961}, thermomass (TM) \cite{BYCao2007}, Dual Phase Lag model \cite{Tzou1995,DYTzou1997}, and the temperature wave theory \cite{Kuang2014}. 
 	 
 	Among the established methods, the dual phase lag model with two time lags of heat flux, $\tau_q$, and temperature gradient, $\tau_t$, leading to engineering analysis with acceptable accuracy, has been recognized as a simple and straightforward macroscopic formulation for the heat transfer study at microscopic levels \cite{Ghazanfarian2015,Shomali2021}.
 	
 	\begin{equation}
 		\vec{q}(\vec{r},t+\tau_q)=-k\vec{\nabla}T(\vec{r},t+\tau_t).
 		\label{ConstitutiveDPL}
 	\end{equation}
 	
 	In the above equation, k, $\vec{q}$, and $\nabla T$, are, respectively, the thermal conductivity, the heat flux vector and the temperature gradient. It has been shown that the DPL model can predict the transient heat propagation behaviors validated through the experiment results \cite{Antaki2005}. It is established that using the DPL model for studying the temperature and heat flux behavior through the nanostructures results in data which are almost near to that of the atomistic methods. Although the DPL results are not as precise as data from atomistic calculations, but they have advantages of being obtained in the blink of an eye. This low computational cost makes DPL a reasonable method in nanoscale.
  
 Although the DPL model introduces the temporally nonlocal effects by implementing the thermal lagging, it passes over the heat nonlocal behavior in space which is an inevitable fact for the heat conduction at nanoscale. Although many studies have focused on the temporal behavior of the heat transfer at the nanoscale, fewer inquiries have been carried out considering the spatial attitude of the non-Fourier heat transfer. Whereas to obtain the detailed information about the heat transfer situation in micro/nano scale, and predicting the thermal performance, that allows to design nano-devices with optimal thermal conditions, it is crucial to include non-locality terms. In principle, the quasi-ballistic heat flux in/at a given location and time intrinsically depends on the temperature gradient in other places, in addition to the earlier times. Further, the constitutive law is no longer localized in a non-diffusive transport regime, but instead possesses memory in space beside the time memory. On the other hand, experimental work has confirmed that the thermal conductivity of micro/nano materials, like silicon nanofilms or nanowires, is size-dependent. This is while the nano-sized structures thermal conductivity differs remarkably from that of the bulk materials \cite{Schelling2002,MXu2019} and nonetheless, the DPL model is not successful in acclimating the size-dependent thermal conductivity.

Several mathematical models have been recommended to account for the spatially nonlocal effects in nanoscale heat conduction \cite{Abouelregal2025}. One widely recognized method is the Guyer-Krumhansl model, which originated from the linearized Boltzmann equation for the pure phonon field \cite{RAGuyer1966,RAGuyer1966-2}. Further, the thermomass model considering the mass, pressure and the phonon gas inertial force is also remarkable \cite{BYCao2007}. Another interesting way to employ the non-locality into heat transport is using the fractional dynamics. Specifically, the fractional temperature equation has been introduced to present a non-local model of thermal energy transport \cite{Mongiovi2013,Shomali2022,Salman2024}. Beside, in analogy with the integral form of the nonlocal thermal conductivity and elasticity, an integro-differential equation that governs nonlocal effects have been proposed by Vermeersch and shakuri \cite{Vermeersch2014}, and Xu \cite{MXu2018,MXu2019}. Evidently, there exists no standard agreement yet on the spatial nonlocality in heat conduction \cite{YJYu2016}. Meanwhile, Tzou and Guo extended Tzou's dual phase lag model to include nonlocal effects in heat flux \cite{DYTzou2010,DYTzou2011}. The so-called Nonlocal DPL model has an advantage of macroscopically taking temporally and spatially nonlocal effects in a simple form. This formalism makes it possible to perform micro/nano thermal engineering analysis without the necessity of having the phonon dynamics knowledge and has also very lower computational cost \cite{WPeng2024,LHai2024,Chawla2024}. In particular, it should be mentioned that the stationary and transient responses of the model are validated experimentally by the size-dependent thermal conductivity of silicon nanofilms and the transient temperature variation throughout the femtosecond laser heating of gold films \cite{WYang2020}. Previously, the Non-local DPL model has been further applied to more precisely study of the temperature and heat flux profile in one-dimensional silicon transistors \cite{Roya2023}. Barati and Shomali have obtained the non-local parameter for one-dimensional silicon transistors and interestingly showed that the parameter is linearly dependent on the Knudsen number. In the present work, we will justify why DPL results are not very accurate and introduce the non-locality as a solution. It is obtained that implementing the non-locality in heat flux is “the missing piece of the puzzle”. While the conventional DPL presents not perfect but almost good results for high Kn number systems, including the non-locality makes the results, also for High Kn numbers, very close to that of the atomistic data.
Bringing the non-locality issue to more realistic cases, in the present study, the two-dimensional silicon transistor is investigated. The importance of heat transport study in transistors comes from the dependency of its reliability on the obtained maximum temperature during the functionality. To put it differently, when the temperature and heat flux profiles are determined, one can trace the thermal behavior inside the single transistor or on a whole die, to use it for the thermal management solutions such as the heat spreader in an appropriate place \cite{Subrina2009,Shomali2012,Shomali20152,Shomali2016,Shomali2017,2Shomali2017,Shomali2023}. Here, in continuation of the previous work, the non-locality coherence length alongside the other unknown adjustable parameters is obtained for two-dimensional silicon transistors via verification of the results with the data available from solving the phonon Boltzmann transport equation (PBTE) \cite{Yang2005}. It is acquired that the dimensionless non-local parameter for the two-dimensional case still has linear dependency related to the Knudsen number but with a coefficient being one-half that of the one-dimensional system. In the present paper, in Sec. \ref{Sec.2}, the researched geometry and the appropriate boundary conditions are presented. Sec. \ref{Sec.3}, introduces the mathematical framework for the two-dimensional non-local dual phase lag model. Then, Sec. \ref{Sec.4} will explain about the numerical concerns. Finally, the results are given in Sec. \ref{Sec.5} and, the conclusion comes in Sec. \ref{Sec.6}.

\section{Geometry and Boundary conditions}
\label{Sec.2}
\begin{figure}
\centering
\vspace{-6mm}
\includegraphics[width=\columnwidth]{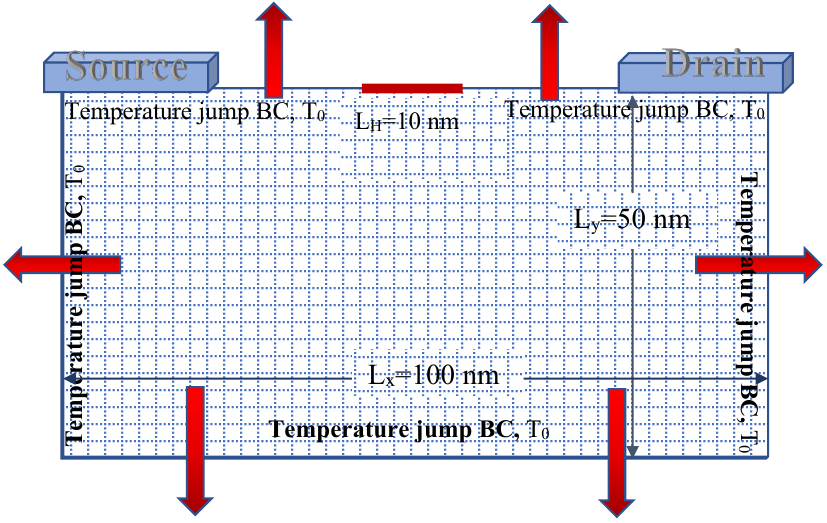}
\caption{\label{geo} Schematic geometry of the two-dimensional silicon MOSFET with 10 nm heater located at the top boundary.}
\end{figure}

In the present research, the transient and steady state behavior of the thermal transport within a simplified two-dimensional silicon metal oxide semiconductor field effect transistor (MOSFET) have been investigated. The geometry which is analogous to the one studied by Yang \emph{et al.} \cite{Yang2005}, is illustrated in Fig. \ref{geo}. Yang and his coauthor \cite{Yang2005} has implied the Boltzmann transport (BT) and the Ballistic-Diffusive (BD) equations for this configuration. Similarly, Ghazanfarian and Shomali have investigated transient heat transfer in a similar 2D nanoscale geometry using a common dual-phase-lag model \cite{Ghazanfarian2012}. The length and width of the two-dimensional transistor are, respectively, 100 and 50 nanometer. As the Fig. \ref{geo} presents, for the 10 nm length on the middle of the top boundary of the 2D silicon, the heater is placed. The heater dimension, L$_H$, is adjustable to control the phonon Knudsen number which is defined as $Kn=\frac{\lambda}{L_h}$ with $\lambda$ being the phonon mean free path. Initially, at time t$^*$=0, the whole system is kept at an ambient temperature T$_0$=300 K, while the heater temperature is suddenly increased to T$_1$=600 K. All the boundaries are exposed to the ambient temperature and, further, a temperature jump boundary condition is applied on all boundaries in order to consider boundary phonon scattering at the micro/nano scale: 

\begin{equation}
	T_s-T_w=-\alpha Kn(\frac{\partial T}{\partial n})_w.
	\label{boundrycondition}
\end{equation}

In the above equation, $T_s$ and $T_w$ are, respectively, the jump temperature of the wall and the boundary temperature. In addition, $\alpha$, is the coefficient which will be rectified to fulfill the boundary condition \cite{Basirat2006,Ghazanfarian2009}. Furthermore, while the index term $w$ encompasses all the boundaries within the solution domain, the parameter $n$, indicates the direction of the outward normal vector at each boundary.

\section{Mathematical Modeling and boundary conditions}
\label{Sec.3}
Heat conduction is arguably the most important dissipative phenomenon and forms the basis of all dissipation theories, especially in the context of non-equilibrium thermodynamics \cite{Zhmakin2023}. In recent decades, several mathematical models have been proposed to explain this phenomenon. The first heat conduction model was introduced by Joseph Fourier in the early 19th century, establishing a phenomenological linear relationship between the heat flux vector and the temperature gradient. However, in addition to the significant dependence of thermophysical properties on size and temperature, heat transfer at the micro/nanoscale exhibits unique properties that cannot be adequately described by Fourier's law, which is based on macroscopic averages and equilibrium concepts. Therefore, the modification of Fourier's law and the development of constitutive models for nanoscale heat conduction to account for nonlocal and memory effects has been proposed. In 1994, Tzou introduced the dual-phase lag (DPL) model which takes into account non-locality in time and in 2010, Tzou and Guo generalized the model by incorporating nonlocal effects in space. The newly non-local DPL model, called here as NDPL, is formulated as follows,

\begin{equation}
\vec{q}(\vec{r} + \vec{\gamma_q} , t+\tau_q) = -K\vec{\nabla}T(\vec{r} , t+\tau_T).
\end{equation}

For two dimensional cases, the equation is written as:

\begin{equation}
	\vec{q}(x + \gamma_{q_x}, y + \gamma_{q_y}, t + \tau_q) = -K\vec{\nabla}T(x,y , t+\tau_T).
	\label{NDPL}
\end{equation}

Here, $\tau_q$ and $\tau_T$, being both positive and intrinsic material properties, are subsequently, the phase lag of the heat flux and the temperature gradient. The $\gamma_q$ is the correlating length of the non-local heat flux. The thermal phase lag values, in analogy to other thermal properties of engineering materials such as the thermal conductivity and diffusivity, are determined theoretically or experimentally under various circumstances. By employing the first-order Taylor series expansion of Equation \ref{NDPL} relative to the non-local coherence length and the phase lags, the non-local DPL heat conduction model is developed:

	\begin{align}
		\vec{q}(x, y, t) &+ \gamma_{q_x} \frac{\partial q(x, y, t)}{\partial x} + \gamma_{q_y} \frac{\partial q(x, y, t)}{\partial y} + \tau_q \frac{\partial \vec{q}(x, y, t)}{\partial t} \nonumber \\
		&= -K \left[ \nabla T(x, y, t) + \tau_T \frac{\partial \nabla T(x, y, t)}{\partial t} \right].
		\label{NDPL1}
	\end{align}

On the other hand, in the absence of an external energy source, the energy equation (\ref{energy eq}) is:

 \begin{equation}
	- \nabla \cdot \vec{q}(r,t) =  C \frac{\partial T(r,t)}{\partial t},
	\label{energy eq}
\end{equation}

with $T$ and $\vec{q}$ being the temperature and the heat flux and $C$ representing the specific heat. By computing the divergence of equation (\ref{NDPL1}) and substituting the expression $ \nabla \cdot \vec{q}$ from equation (\ref{energy eq}) into it, one obtains the following equation for the temperature:

\begin{equation}
	\label{13}
	\begin{aligned}
		& \frac{K}{C \tau_q}\left(\frac{\partial^2 T}{\partial x^2}+\frac{\partial^2 T}{\partial y^2}\right)+\frac{K \tau_T}{C \tau_q}\left(\frac{\partial^3 T}{\partial x^2 \partial t}+\frac{\partial^3 T}{\partial y^2 \partial t}\right) \\
		&= \left(\frac{1}{\tau_q} \frac{\partial T}{\partial t}+\frac{\vec{\gamma_q}}{\tau_q}\left(\frac{\partial^2 T}{\partial x \partial t}+\frac{\partial^2 T}{\partial y \partial t}\right) \right)+ \frac{\partial^2 T}{\partial t^2}.
	\end{aligned}
\end{equation}

Also, the equation for calculating the heat flux, $\vec{q}$, becomes as:
\begin{equation}
			\label{14}
   \begin{aligned}
	\vec{q}(x, y, t)+&\gamma_{q_x} \frac{\partial q(x, y, t)}{\partial x} + \gamma_{q_y} \frac{\partial q(x, y, t)}{\partial y} + \tau_q \frac{\partial \vec{q}(x, y, t)}{\partial t} \\
	&= -K \left[ \nabla T(x, y, t) + \tau_T \frac{\partial \nabla T(x, y, t)}{\partial t} \right].
	\end{aligned}
\end{equation}

Now the equations are obtained, it is better to deal with the boundary conditions. Phonon scattering plays a crucial role in heat transport in nanomaterials especially at the boundary surfaces. The study by Asheghi \emph{et al.} \cite{Asheghi1997} has demonstrated that the phonon scattering at the layer boundaries significantly reduces the thermal conductivity in thin silicon layers. In a related work, by taking into account the phonon scattering and consequently applying a jump boundary condition at the semiconductor-oxide interface, a significant reduction in peak temperature is reported  \cite{Jiaung2008}. So, dealing with the present study, applying no-jump boundary conditions in our calculations leads to unsatisfactory results especially near the boundaries. Therefore, we use the same mixed-type boundary condition that Ghazanfarian and Abbassi considered for boundary phonon scattering \cite{Ghazanfarian2009,Ghazanfarian2012-2}. They have shown, numerically and analytically, that using the mixed boundary condition with the DPL model, sufficiently captures the boundary temperature jump \cite{Shomali2012}. In other words, using the mixed-type boundary condition is essential to model the heat transfer at nanoscale accurately.

\subsection{normalization}
In order to eliminate the complexity and further, normalization of Eqs. \ref{13}, \ref{14}, and also the temperatrue jump boundary condition, \ref{boundrycondition}, the following non-dimensional parameters are introduced:
\begin{equation}
	\nonumber
	\begin{aligned}
	\vec{\gamma*} &= \frac{\vec{\gamma_q}}{L_H}, \quad T^* = \frac{T - T_0}{T_0}, \quad q^* = \frac{q}{C \left| v \right| T_0} \quad K = \frac{c v \lambda}{3}, \\
		K n &= \frac{\lambda}{L_H}, \quad X = \frac{x}{L_x}, \quad Y = \frac{y}{L_y}, \quad x^* = \frac{x}{L_H}, \quad y^*= \frac{y}{L_H}, 
	\end{aligned}
		\label{dimensionless}
\end{equation}
\begin{equation}
			 B=\frac{\tau_t}{\tau_q}, \quad t^* = \frac{t}{\tau_q} 
	\end{equation}

 In the above formulas, the notation $*$, denotes the dimensionless condition. Besides, $C$, $k$, $v$ and $\lambda$ are, respectively, the specific heat, the thermal conductivity, the group velocity of the phonon, and the mean free path of the phonon. Also, the parameter $L_h$, is the heater length of the transistor. By substituting the dimensionless quantities from Eq. \ref{dimensionless} into Eq. \ref{13}, one obtains the following dimensionless temperature equation:

\begin{equation}
	\begin{aligned}
		\label{15}
		&\frac{\partial T^*}{\partial t^*}+\vec{\gamma} \frac{\partial}{\partial t^*}\left(\frac{\partial T^*}{\partial x^*}+\frac{\partial T^*}{\partial y^*}\right)+\frac{\partial^2 {T}^*}{\partial t^{* 2}} \\
		& \left(=\frac{Kn^2}{3}\left(\frac{\partial^2 {T}^*}{\partial {x}^{* 2}}+\frac{\partial^2 {T}^*}{\partial y^{* 2}}\right)+\frac{B {Kn}^2}{3} \frac{\partial}{\partial t^*}\left(\frac{\partial^2 {T}^*}{\partial {x}^{* 2}}+\frac{\partial^2 {T}^*}{\partial {y}^{* 2}}\right)\right)
	\end{aligned}
\end{equation}

Similarly, contemplating the above non-dimensional parameters in the equation for energy, Eq \ref{14}, results in: 

 \begin{equation}
 		\begin{aligned}
	\vec{q^*} + &\vec{\gamma*} \left[ \frac{\partial q^*}{\partial x^*} + \frac{\partial q^*}{\partial y^*}  \right] + \frac{\partial q^*}{\partial t^*} 
	\\
	&= - \frac{kn}{3} \left[\left(\frac{\partial T^*}{\partial x^*} + \frac{\partial T^*}{\partial y^*}\right) + B \frac{\partial}{\partial t^*} \left(\frac{\partial T^*}{\partial x^*} + \frac{\partial T^*}{\partial y^*}\right)\right].
	\label{16}
		\end{aligned}
\end{equation}

Finally, the temperature jump boundary, Eq. \ref{boundrycondition}, is modified into the equation,

\begin{equation}
T^*_s-T^*_w=-\alpha Kn(\frac{\partial T^*}{\partial n^*})_w.
\end{equation}

Here, $n^*$ is the dimensionless boundary normal vector. Two unknown parameters, $B$ and $\gamma$, alongside the parameter $\alpha$ are determined in order to make the results of the NDPL model compatible with the data obtained from solving the phonon Boltzmann transport equation. In more detail, our results obtained through the NDPL model are compared with the findings from Yang \emph{et al.} \cite{Yang2005}. This comparison provides a benchmark for evaluating the reliability and accuracy of our simulation outcomes.

\section{Numerical considerations}
\label{Sec.4}
Since Eqs. \ref{15} and \ref{16} are, respectively, a third and second-order partial differential equations with the mixed derivatives in $t^*$, $x^*$ or $y^*$, special attention is required to avoid the convergence and stability problems in the numerical discretization. Thus, a stable and convergent three-stage finite difference scheme, is proposed by Dai \emph{et al.} \cite{Dai2004} in 2004 for the one-dimensional DPL model in spherical coordinates. This method uses a weighted average for ensuring the stability and the uniformity. Ghazanfarian and Shomali have adopted the 2D scenario in Cartesian coordinates \cite{Ghazanfarian2012}. Here, the same is utilized. More particularly, the eqs. \ref{15} and \ref{16}, are solved numerically using a fully implicit second-order finite difference method with central discretization of all derivatives. For instance, the discretization of the terms in the left-hand side of the Eq. \ref{15} are expressed in the following way,

\begin{equation}
	\frac{\partial T^*}{\partial t^*}=\frac{T_{i, j}^{* n+1}-T_{i, j}^{* n-1}}{2 \Delta t},
\end{equation}
\begin{equation}
	\frac{\partial^2 T^*}{\partial t^* \partial x^*} = 
	\frac{1}{4 \Delta x \Delta t} \left[ \left( T_{i+1, j}^{* n+1} - T_{i-1, j}^{* n+1} \right) - \left( T_{i+1, j}^{* n-1} - T_{i-1, j}^{* n-1} \right) \right],	
\end{equation}
\begin{equation}
	\frac{\partial^2 T^*}{\partial t^* \partial y^*} = 
	\frac{1}{4 \Delta y \Delta t} \left[ \left( T_{i, j+1}^{* n+1} - T_{i, j-1}^{* n+1} \right) - \left( T_{i, j+1}^{* n-1} - T_{i, j-1}^{* n-1} \right) \right],
\end{equation}
\begin{equation}
	\frac{\partial^2 {T}^*}{\partial t^{* 2}}=\frac{T_{i, j}^{* n+1}+2 T_{i, j}^{* n}+T_{i, j}^{* n-1}}{\Delta t^2},
\end{equation}

and the right side of the Eq. \ref{15} is discretized as:

\begin{equation}
\frac{Kn^2}{3}[P_x(T^*_{av}) + P_y(T^*_{av}) + B{[P_x(T^*_d) + P_y(T^*_d)}].
\end{equation}

where, the operators $P_x$ and $P_y$, in the above relation are defined as:

\begin{equation}
	\begin{aligned}
		& P_y\left(T_{i, j}^*\right)=\frac{T_{i, j+1}^*-2 T_{i, j}^*+T_{i, j-1}^*}{\Delta y^2}, \\
		& P_x\left(T_{i, j}^*\right)=\frac{T_{i+1, j}^*-2 T_{i, j}^*+T_{i-1, j}^*}{\Delta x^2}, \\
		& T_{a v}^*=\frac{T_{i, j}^{* n+1}+2 T_{i, j}^{* n}+T_{i, j}^{* n-1}}{4}, \\
		& T_d^*=\frac{T_{i, j}^{* n+1}-T_{i, j}^{* n-1}}{2 \Delta t}.
	\end{aligned}
\end{equation}

So, the whole Eq. \ref{15} is written as:

\begin{equation}
	\nonumber\\
	\centering
	 \frac{T_{i, j}^{* n+1}-T_{i, j}^{* n-1}}{2 \Delta t}+\vec{\gamma}(	\frac{1}{4 \Delta x \Delta t} \left[ \left( T_{i+1, j}^{* n+1} - T_{i-1, j}^{* n+1} \right)- \left( T_{i+1, j}^{* n-1} - T_{i-1, j}^{* n-1} \right) \right]+
	 \end{equation}
	 \begin{equation}
	 		 	\centering
	 \frac{1}{4 \Delta y \Delta t} \left[\left( T_{i, j+1}^{* n+1} - T_{i, j-1}^{* n+1} \right)- \left( T_{i, j+1}^{* n-1} - T_{i, j-1}^{* n-1} \right) \right])
	 \end{equation}
	 \begin{equation}
	 				 	\nonumber
	+\frac{T_{i, j}^{* n+1}+2 T_{i, j}^{* n}+T_{i, j}^{* n-1}}{\Delta t^2} =\frac{Kn^2}{3}[P_x(T^*_{av}) + P_y(T^*_{av}) + 
	\end{equation}
	\begin{equation}
			 	\nonumber
	B{[P_x(T^*_d) + P_y(T^*_d)}].
	\end{equation}

It is worthwhile to note that a uniform 301$\times$151 numerical mesh with the dimensionless time-step size of 10$^{-3}$ is found to be adequate for pursuing a mesh and time step size independence study.
\section{Results and Discussions}
\label{Sec.5}
 
\begin{figure*}[h!]
	\centering
	\includegraphics[width=1.7\columnwidth]{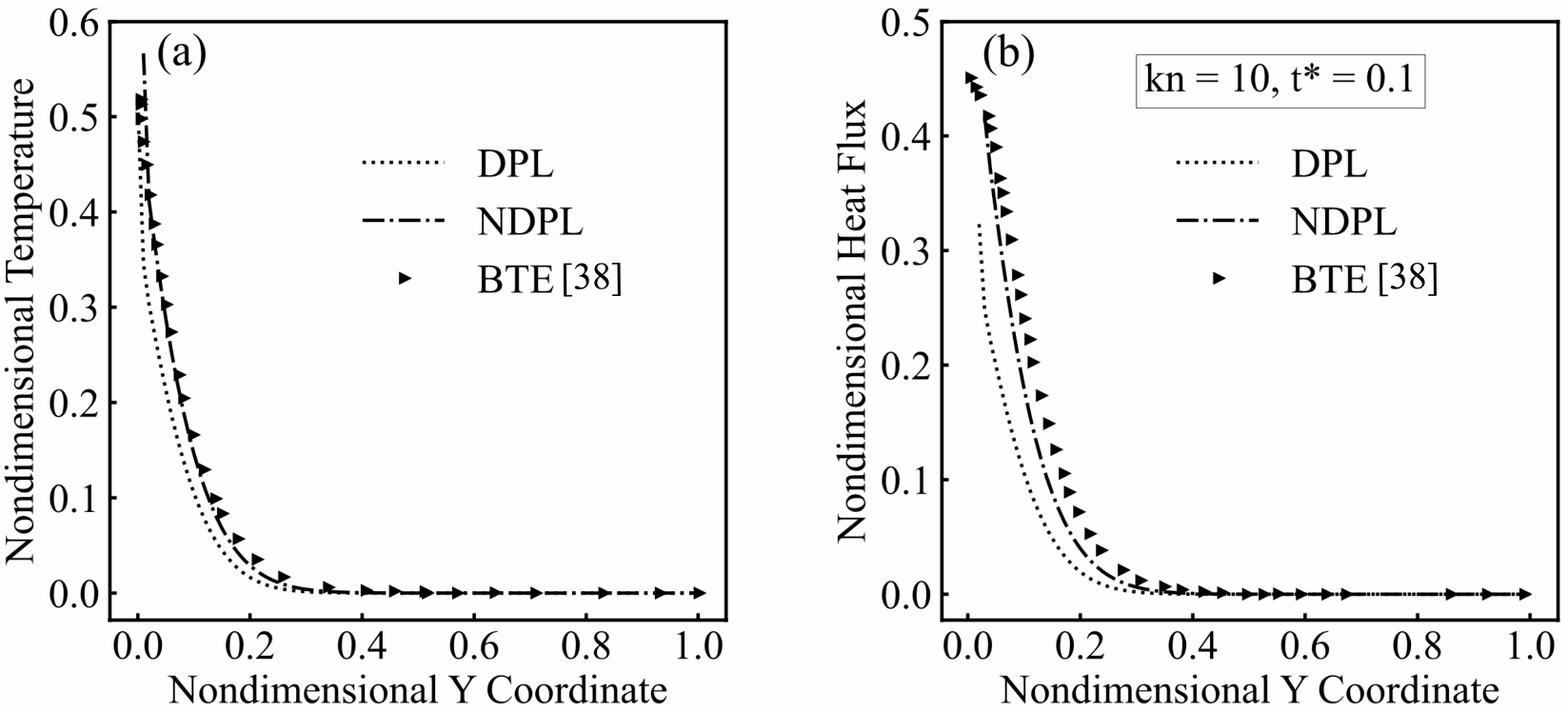}
	\caption{\label{kn10t0.12D} (a) The temperature and (b) the heat flux profile calculated from the NDPL, DPL, and phonon BTE when Kn=10 and t$^*$=0.1. The profiles are shown at centerline where X=0.5.}
\end{figure*}

In this section, the heat transfer simulation results for a simplified 2D MOSFET using the non-local dual phase lag model are presented. The temperature and heat flux profiles which are obtained from the nonlocal DPL and the standard DPL, are shown in non-dimensional form. To verify our findings, we have conducted a comparative analysis of our results with those obtained by Yang \emph{et al.} \cite{Yang2005}. In addition, the acquired results are compared with the outcomes generated by Ghazanfarian and Shomali \cite{Ghazanfarian2012}. They have modeled the temperature and heat flux profiles of two-dimensional MOSFETs for various Knudsen numbers over different time periods using the standard DPL framework. It has been pointed out that as the Knudsen number increases, there appears a notable divergence between the obtained profiles, from the ones predicted by the PBE. This discrepancy, also announced in one-dimensional results \cite{Ghazanfarian2009}, has been previously attributed to the ignorance of the non-locality in space in DPL model \cite{Roya2023}.

The objective of this study is to address the discrepancies observed at high Knudsen numbers by employing a nonlocal heat transport model. By introducing a non-dimensional correlation length of $\gamma$=1.5, this study highlights the crucial role of nonlocal effects in accurately modeling heat flux in nanostructures at large Knudsen numbers.
To elucidate further, Fig. \ref{kn10t0.12D} illustrates the enhanced precision of the outcomes derived from the Nonlocal Dual-Phase-Lag (NDPL) model in comparison to the conventional DPL model for a two-dimensional transistor with a Knudsen number of 10. Taking into account B=0.08 and $\alpha$=0.02, the Fig. \ref{kn10t0.12D} (a) illustrates the temperature profile for a Knudsen number of 10 at centerline, where X=0.5. The temperature distributions derived from the NDPL model for scaled times of t$^*$=0.1 and t$^*$=1 closely resemble solutions derived from the phonon Boltzmann transport equation. It is notable that, in this case, the NDPL model consistently aligns more closely with BTE results than the standard Dual-Phase-Lag model.
The NDPL model exhibits precise alignment with the BTE outcomes for scaled coordinates below X=0.8 at t$^*$=1 and across all positions at t$^*$=0.1. At t$^*$=0.1, the system displays pronounced non-Fourier behavior, attributable to the relatively small size and time scales involved. Although the system continues to exhibit non-Fourier behavior at t$^*$=1, the deportment is less pronounced than that of the t$^*$=0.1. Fig. \ref{kn10t0.12D} (a) demonstrates that incorporating nonlocality into the DPL model effectively addresses non-Fourier behavior across nearly all positions, resulting in temperature distributions that closely align with the BTE results. At larger times, such as t$^*$=10, the impact of incorporating nonlocality diminishes as the system reaches a steady state, and non-Fourier behavior becomes less influential.
The Fig. \ref{kn10t0.12D} (b) illustrates the relationship between the heat flux and position. Furthermore, the superior reliability of the NDPL modeling approach, as compared to the standard DPL approach with the phonon BTE data set, is evident. The consistency of the output data of NDPL modeling with the available data from BTE simulation, in comparison to the results of the standard DPL, is further noticeable. In other words, at t$^*$ = 0.1, the heat flux aligns closely with the BTE results, exhibiting a behavior similar to the temperature profile. In particular, for high Knudsen numbers, the heat flux distribution derived from DPL modelling is found to deviate significantly from the computed temperature profile, resulting in a notable discrepancy between the predicted and actual outcomes of PBE. Interestingly, incorporating nonlocality into the DPL model effectively resolves this issue, leading to more accurate solutions. For longer time scales, the role of nonlocality and lagging behaviour becomes less pronounced, with $\gamma$ and $B$ approaching zero. The anticipated outcome is due to the transistors eventually reaching a steady state, thus reducing the impact of non-Fourier heat transfer. Fig. \ref{kn10t0.12D} indicates that the temperature jump coefficient $\alpha$ is the key parameter for achieving accurate temperature and heat flux profiles. In this study, the value of $\alpha$ is adjusted based on the Knudsen number over longer timescales in order to correct discrepancies in the heat flux profiles obtained from the standard DPL model.

\begin{figure*}[h!]
	\centering
	\includegraphics[width=1.3\columnwidth]{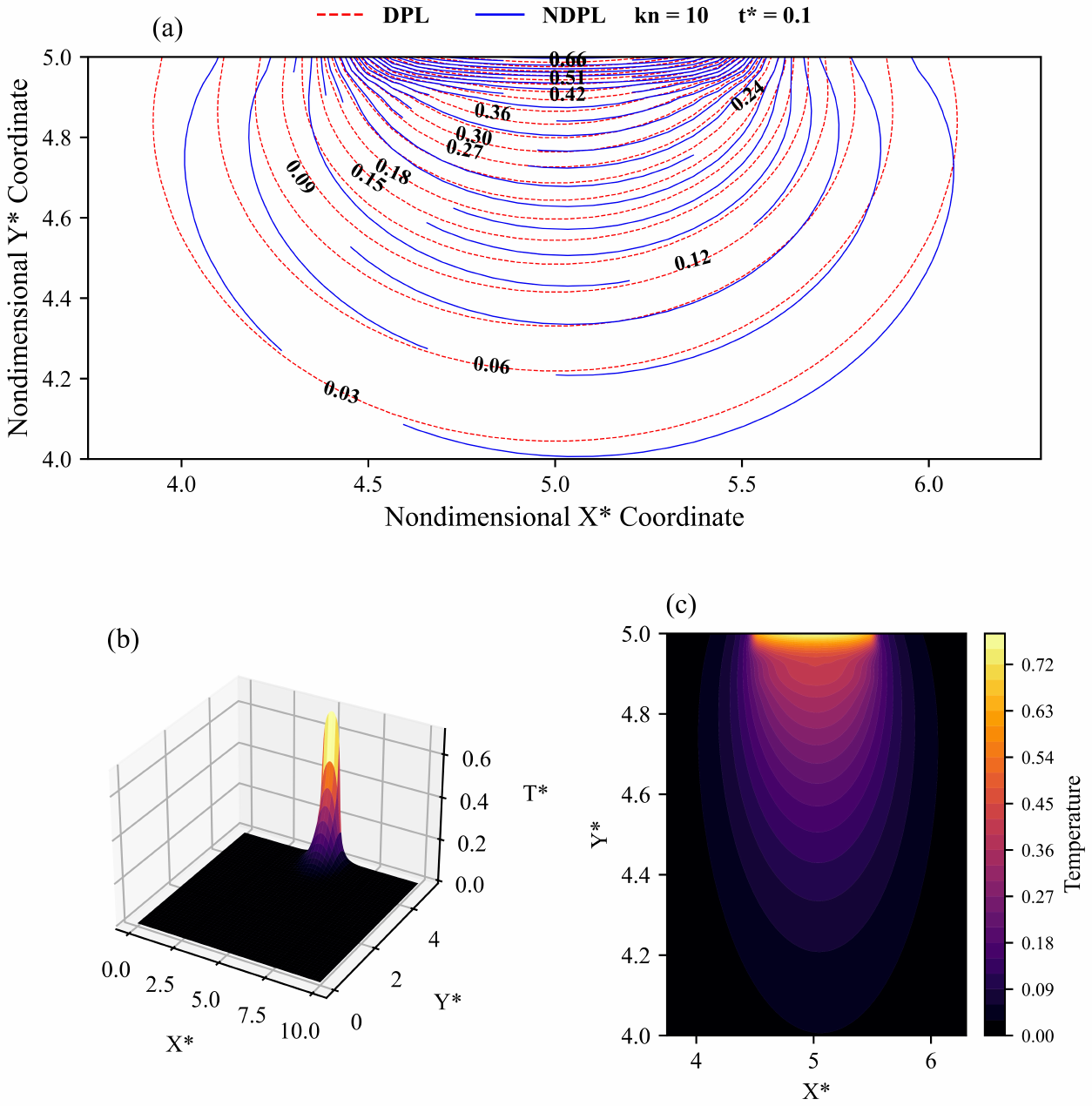}
	\caption{\label{cont0110} Comparison of the temperature distribution obtained from the Boltzmann Equation \cite{Yang2005}, right triangle, the nonlocal behavior DPL model, dashed line, and the standard DPL model for different instantaneous times and Kn=10 at t$^*$=0.1.}
\end{figure*}

 Moreover, the Figs. \ref{cont0110} (a-c) show the contour and surface plots of heat prorogation in the device with Kn=10 at t$^*$=0.1. Particularly, in Fig. \ref{cont0110} (a), where the isothermal diagrams are drawn, the existence of non-locality traces are detectable in asymmetric behavior of the temperature profiles. The non-dimensional non-locality coefficient in both directions is the same, but as the length of the transistor in x dimension is twice that of the y-direction, it seems that the temperature profiles are pushed toward the left side. This means that the effect of non-locality is accumulated as going forward in x- and y- direction. Also, Fig. \ref{cont0110} (b) manifests the behavior of the temperature profile during the first 3.33 ps. As Fig. \ref{cont0110} (b) suggests, at t$^*$=0.1, the heat at the hotter zone affected by the heater, has not had enough time to dissipate all over the transistor. Reaching the temperature 540 K at the hottest zone, it propagates toward the x- and y- directions from both sides leaving most of the layer untouched. Furthermore, the heat propagation is more obvious in Fig. \ref{cont0110} (c), in which the thermal status is determined at every position. It should be mentioned that the Figs. \ref{cont0110} (a) and (b) plots are presented for 4<x*<5 and 4<y*<6, giving emphasize to heater influenced regions. 

\begin{figure}[h!]
	\centering
	\includegraphics[width=0.75\columnwidth]{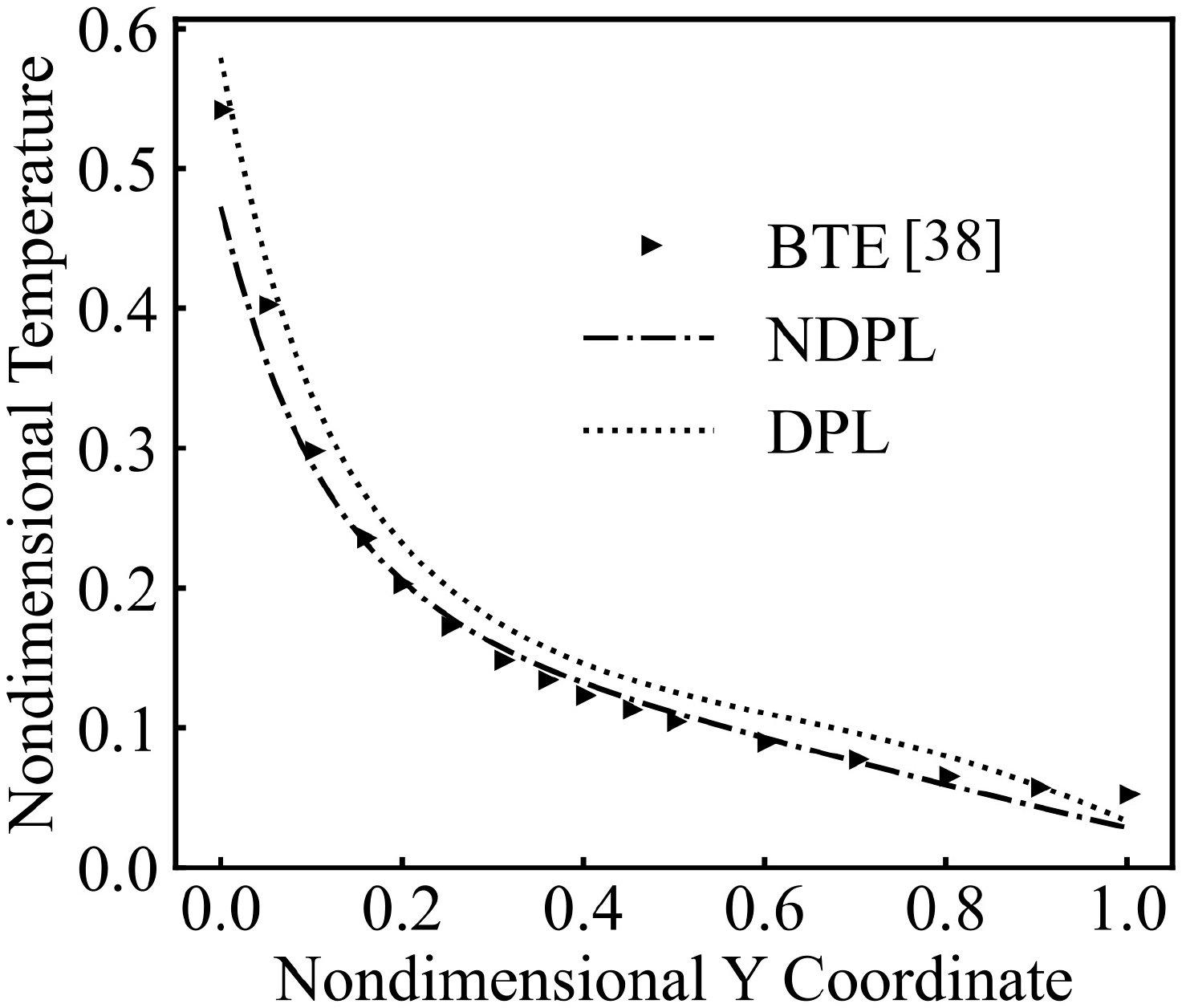}
	\caption{\label{kn10t12D} The temperature profile at the centerline of the geometry, X=0.5, calculated from the NDPL, DPL, and BTE \cite{Yang2005}, when Kn=10 and t$^*$=1.}
\end{figure}

In the next step, the behavior of the temperature profile considering the nonlocal effects have been investigated for the system with Kn=10 at t=33.33 or t$^*$=1. Taking into account B=0.08, the same as the one for t$^*$=0.1, and the parameter $\alpha$ to be 0.1, the result is presented in Fig. \ref{kn10t12D}. The plot shows the superior consistency of the data obtained from NDPL modeling with BTE available data in comparison to the results calculated from the common DPL. This compatibility is such that in the interval x$^*$=0.04 to x$^*$=0.84, the plots obtained from BTE and NDPL are almost the same. Giving consideration to the Figs. \ref{kn10t0.12D} and \ref{kn10t12D}, it is obvious that the implementation of non-locality is the missing piece of the DPL model accurateness puzzle.

\begin{figure*}[h!]
	\centering
	\includegraphics[width=1.3\columnwidth]{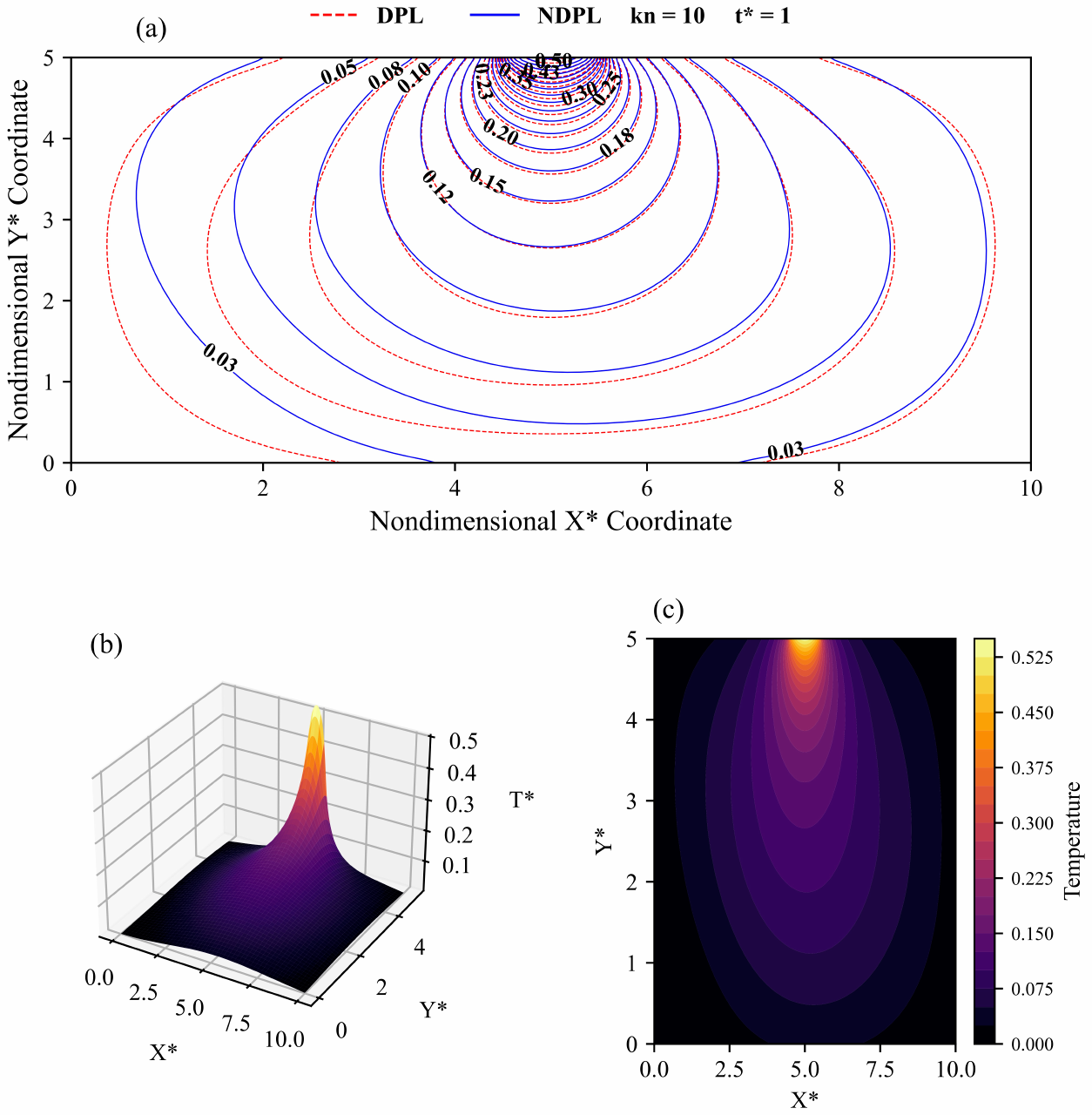}
	\caption{\label{cont10t1} Comparison of the temperature distribution obtained from the Boltzmann Equation \cite{Yang2005}, right triangle, the nonlocal behavior DPL model, dashed line, and the standard DPL model for different instantaneous times and Kn=10 at t$^*$=1.}
\end{figure*}

It is also helpful to look at the thermal behavior of the 2D nano-device at x-y plane. The plot Fig. \ref{cont10t1} (a), confirming the non-local behavior, presents the all-round heat propagation. The size of the layer in x-dimension, is two times the y-dimension. So, the asymmetric behavior is more pronounced at the x-direction due to the more accumulation of the non-locality effect. Furthermore, the Fig. \ref{cont10t1} (b) in comparison to the Fig. \ref{cont0110} (b), presents the propagation of heat flux from the heater at the top boundary to all over the 2D layer. The silicon is thermally isotropic and the mild non-symmetry behavior is attributed to the geometry of the device and existence of the non-locality parameters at x and y directions. The Fig. \ref{cont10t1} (c) also demonstrates, as the heat flux flows, the reached maximum temperature decreases. At the time t=33.33 ps, the system still is in transient regime and the heat flux has not achieved a steady state. One should note that if the material is inherently anisotropic, the value of the non-dimensional non-locality parameter, $\gamma$, at each x and y direction will be different.

\begin{figure}[h!]
	\centering
	\includegraphics[width=0.75\columnwidth]{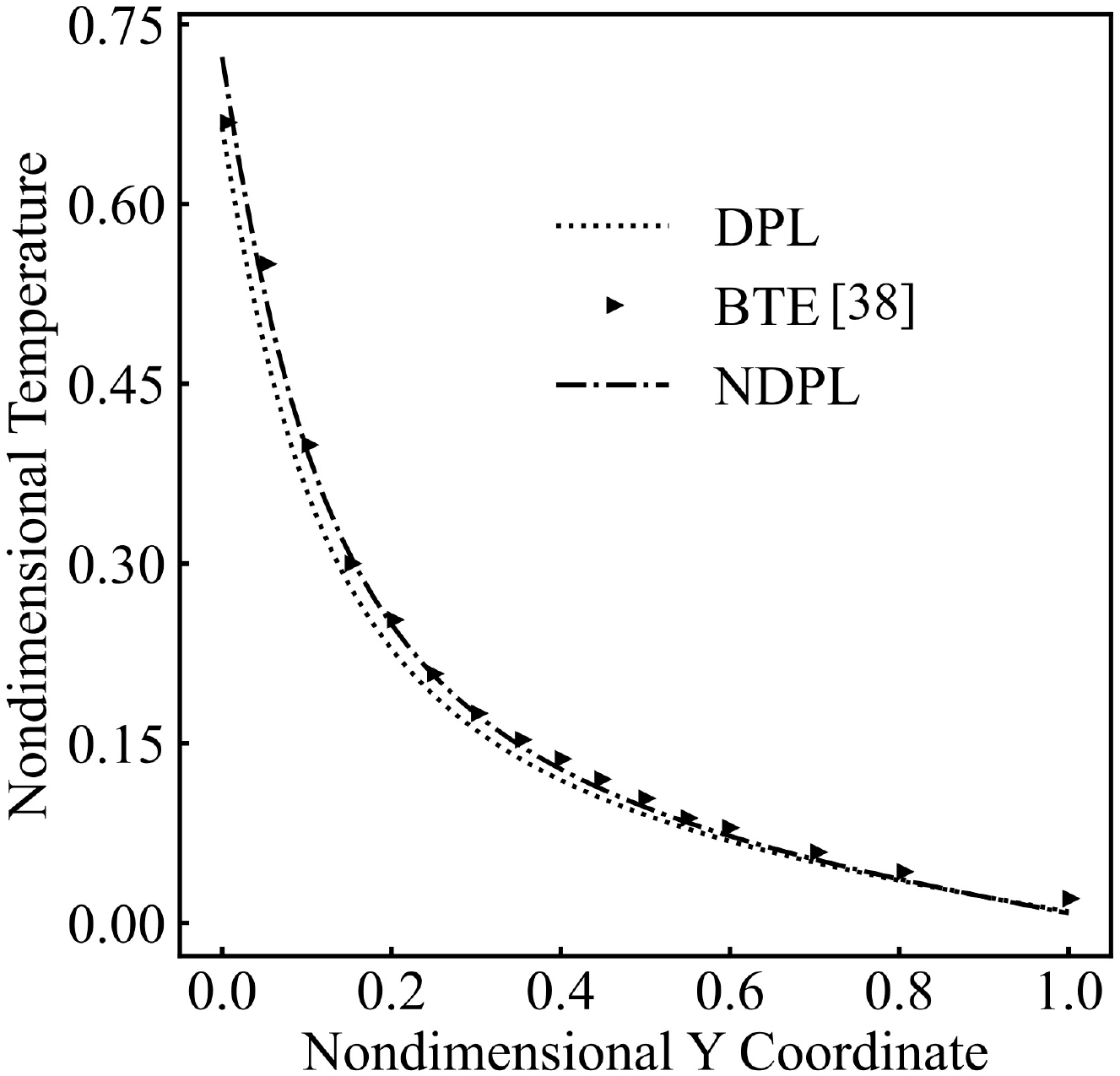}
	\caption{\label{kn1t100} The temperature profile for a nanodevice with Kn=1 at t$^*$=100, calculated from solving the NDPL, DPL, and PB equations \cite{Yang2005}.}
\end{figure}

At the next step, the 2D transistor with Kn=1 being one-tenth of the previous studied structure, is studied. As the Knudsen number decreases, one reaches the macroscopic regime and hence the non-Fourier effects and thus the impact of non-locality in space is expected to fade. The non-dimensional nonlocality parameter, $\gamma$, follows a linear relation relative to the Knudsen number. More precisely, $\gamma_x$ and $\gamma_y$ are equal to 0.15Kn. So, it is set to 0.15 for a system with Kn=1, which is one-tenth of the value for Kn=10. The Fig. \ref{kn1t100} confirms that taking into account B=0.08, $\gamma_x$=$\gamma_y$=0.15, and $\alpha$=0.3, the calculated results from NDPL are in close agreement with those obtained solving the phonon Boltzmann equation \cite{Yang2005}. In essence, the incorporation of nonlocality compensates for nearly all discrepancies between the standard DPL results and the phonon BTE outputs. Further, while the common DPL underestimates the thermal behavior, the non-local DPL model predicts the results precisely. 

\begin{figure}[h!]
	\centering
	\includegraphics[width=0.9\columnwidth]{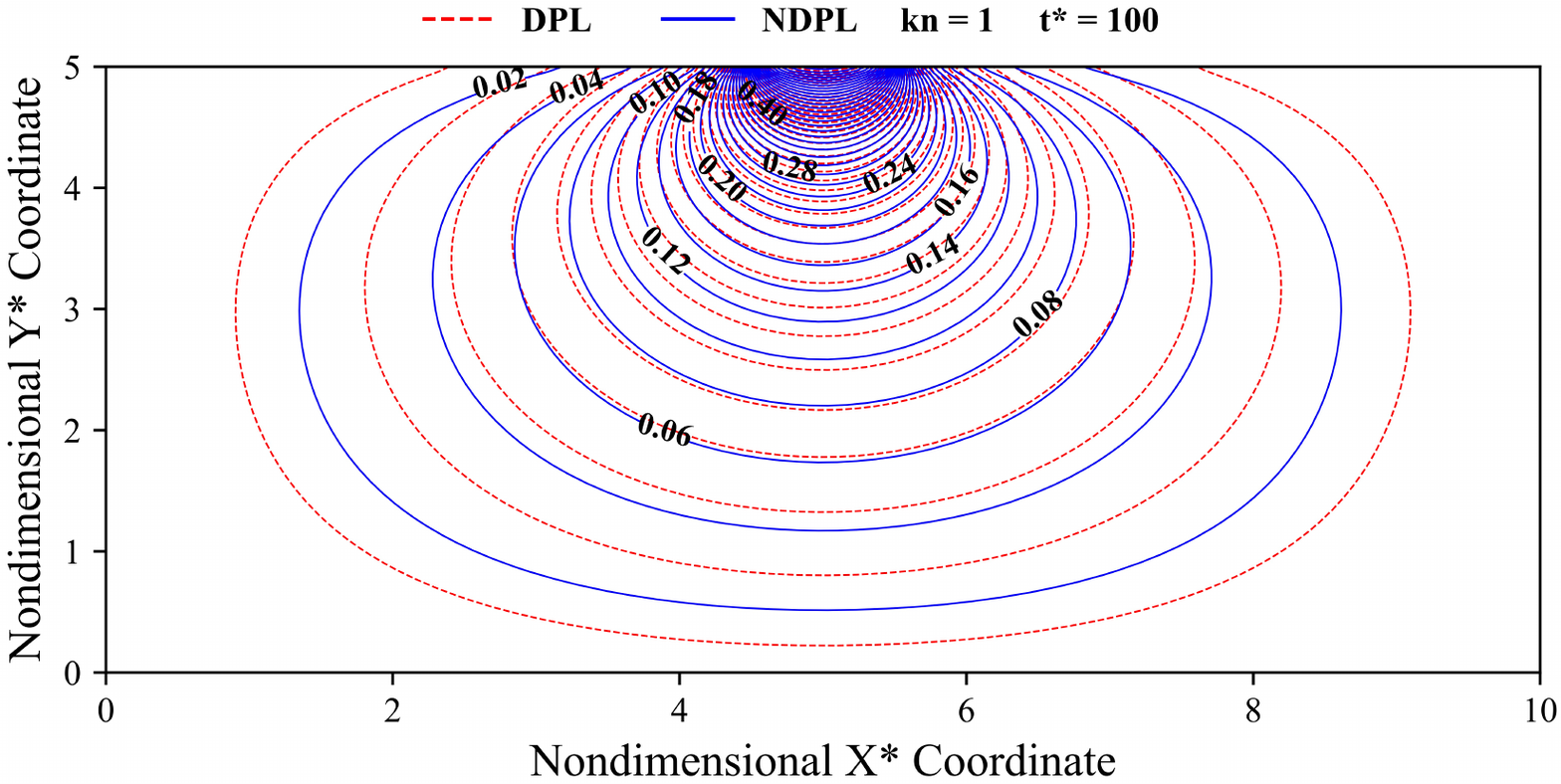}
	\caption{\label{cont1t100} The dimensionless temperature contour plot obtained from the standard and non-local DPL model, respectively, shown as the dashed and solid line, for Kn=1 at t$^*$=100.}
\end{figure}

In Fig. \ref{cont1t100}, the contour plot at time t$^*$=100 are seen. As one expects, when the size is increased and the system presents slighter non-Fourier behavior, and so the non-locality effect becomes less prominent and consequently the non-dimensional non-local parameter gets smaller. Accordingly, as the Fig. \ref{cont1t100} shows the asymmetric behavior due to the existence of the non-locality in heat flux tends to the symmetric one for the system with Kn=1. Also, the difference between the obtained results from the two models confirms the importance of taking non-locality into account.

\begin{figure}[h!]
\centering
\includegraphics[width=0.75\columnwidth]{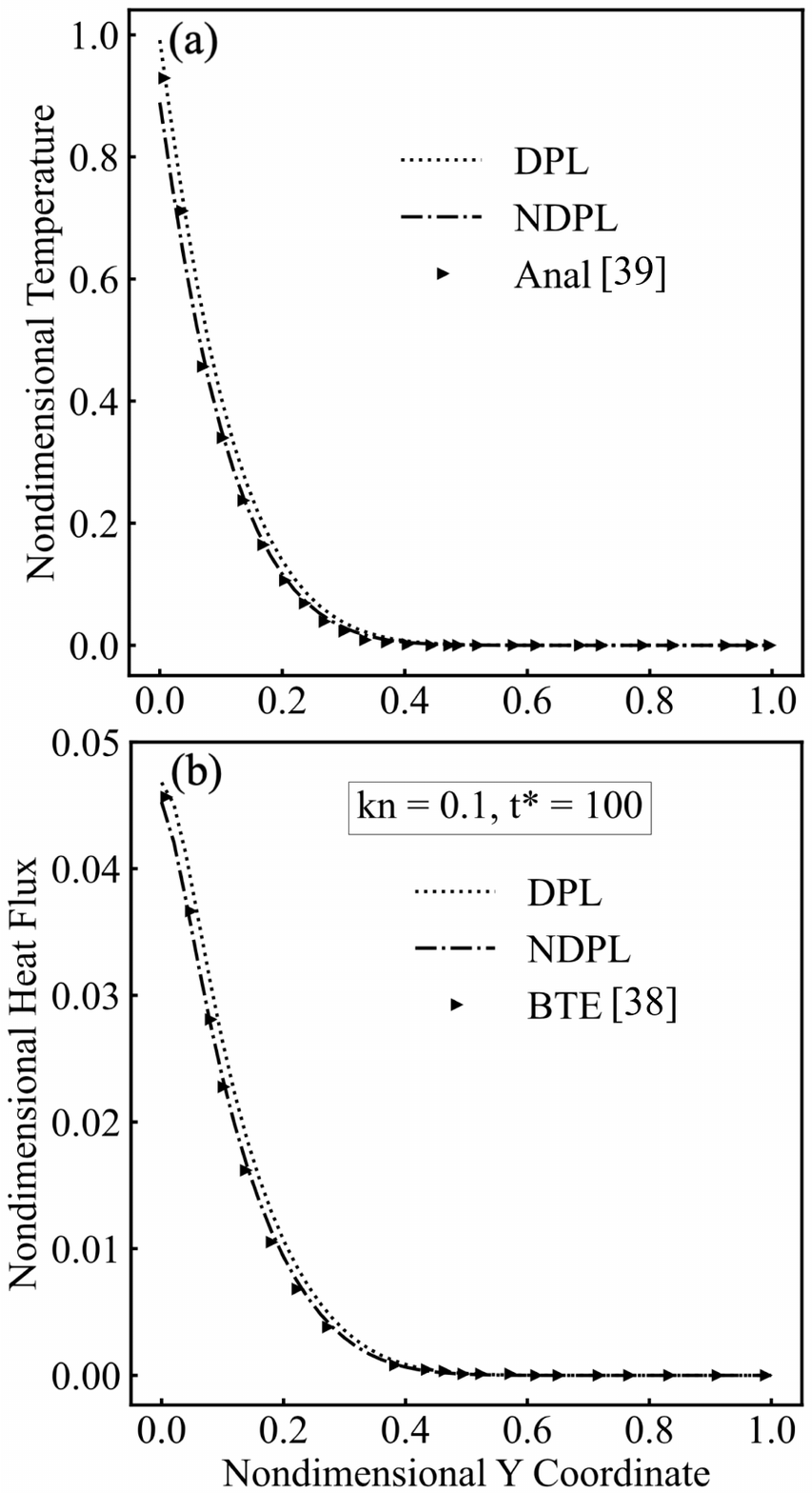}
\caption{\label{kn01t100} (a) The temperature and (b) the heat flux obtained from the phonon Boltzmann transport equation/analytical solution, reported in \cite{Yang2005} and \cite{Ghazanfarian2012-2}, the standard and non-local DPL model, presented respectively by the right triangle, dotted and dashed line, for the system with Kn=0.1 at t$^*$=100. The results are manifested at centerline where X=0.5.}
 \end{figure}

At the final stage, the temperature and heat flux distributions for the system with the low Knudsen number of Kn=0.1 is investigated. Although the value of $\gamma$ parameter for both y and x directions is found to be 0.015, two orders of magnitude smaller than that of the Kn=10, but as the Figs. \ref{kn01t100} (a) and (b) demonstrate, the small gap between the standard DPL result and the available data obtained from analytical calculation \cite{Ghazanfarian2012} and also the PBT equation \cite{Yang2005}, can be covered using the non-local DPL model. Hence, with considering the parameters B and $\alpha$ to be, subsequently, 0.08 and 0.8, one can accurately predict the thermal behavior in Kn=0.1 nano-devices. Inevitably, it should be noted that taking into consideration the non-locality in space, is also crucial for attaining results which are much closer to the realistic condition.  

\section{Conclusion}
\label{Sec.6}

The present work establishes a 2-D framework for micro/nanoscale steady and transient heat transport study to obtain the results which are as precise as that of the atomistic methods and simultaneously need lower computational costs. In particular, the 2-D nonlocal DPL model is introduced with considering the non-locality in space for heat flux alongside the phase lagging, and also the temperature jump boundary condition. More precisely, a non-dimensional non-locality parameter $\gamma$, is implemented to find out the strength of the nonlocality. Once the scheme is formulated, it is utilized for a 2-D silicon MOSFET. The unknown parameters of non-dimensional parameter $\gamma$, in addition to the $\alpha$, the temperature jump coefficient, and B, the phase lagging ratio are newly calculated by verifying the obtained results with the available data from PBE. The $\gamma$ parameter is found to be essential for obtaining the accurate temperature and heat flux profiles. Especially, for the large Knudsen number, the effect of nonlocality at heat flux is very remarkable. Moreover, for Kn=10, the $\gamma$ parameter is found to be 1.5, while the Kn=0.1 system has the $\gamma$ value of 0.015. Furthermore, it is obtained while the silicon is thermally isotropic, the non-symmetric behavior appears in temperature contour plots which is attributed to the device geometry and also the existence of non-locality parameters at x and y directions.

\section*{Acknowledgment} 
This work is based upon research funded by Iran National Science Foundation (INSF) under project No.4027456.


\begin{thebibliography} {10}
	
\bibitem{Samian2013}
R. S. Samian, A. Abbassi, J. Ghazanfarian, Thermal investigation of common 2d FETs and new generation of 3-d FETs using Boltzmann transport equation in nanoscale, Int. J. Mod. Phys. C, 24 (2013) 1350064.

\bibitem{Samian2014}
R. S. Samian, A. Abbassi, J. Ghazanfarian, Transient conduction simulation of a nano-scale hotspot using finite volume lattice Boltzmann method, Int. J. Mod. Phys. C, 25(04) (2014) 1350103.

\bibitem{Moghadam2014}
M. Moghaddam, J. Ghazanfarian, A. Abbassi,Implementation of DPL-DD model for the simulation of nanoscale MOS devices, IEEE Transactions on Electron Devices, 61(9) (2014) 3131.	
	
\bibitem{Shomali2019}
J. Ghazanfarian, Z. Shomali, S. Xiong, 21st Century Nanoscience–A Handbook: Nanophysics Sourcebook (Volume One), Sattler, K. D. (Ed.), Chapter 4, CRC Press, 2019.

\bibitem{Shomali2018}
Z. Shomali, R. Asgari, Effects of low-dimensional material channels on energy consumption of nano-devices,  Int. Commun. Heat Mass Transf., 94 (2018) 77.

\bibitem{Bahadori2024}
S. A. Bahadori, and Z. Shomali, Thermal transport in thermoelectric materials of SnSSe and SnS2: A non-equilibrium Monte-Carlo simulation of Boltzmann transport equation. Case Studies in Thermal Engineering, 57 (2024) 104377.
	
\bibitem{Cattaneo1958}
C. Cattaneo, A form of heat-conduction equations which eliminates the paradox of instantaneous propagation, Comptes Rendus, 247 (1958) 431.

\bibitem{Vernotte1961}
P. Vernotte, Some possible complications in the phenomena of thermal conduction, Comptes Rendus, 252 (1961) 2190.

\bibitem{BYCao2007}
B. Y. Cao, Z.Y. Guo, Equation of motion of a phonon gas and non-Fourier heat conduction, J. Appl. Phys., 102 (2007) 053503.

\bibitem{Tzou1995}
D. Y. Tzou, The generalized lagging response in small-scale and high-rate heating, Int. J. Heat Mass Transf., 38 (1995) 3231.

\bibitem{DYTzou1997}
D. Y. Tzou, Macro- to Microscale Heat Transfer: The Lagging Behavior, Taylor \& Francis, Washington, D.C., USA, 1997.

\bibitem{Kuang2014}
Z. B. Kuang, Discussions on the temperature wave equation, Int. J. Heat Mass Transf., 71 (2014) 424.

\bibitem{Ghazanfarian2015}
J. Ghazanfarian, Z. Shomali, A. Abbassi, Macro-to nanoscale heat and mass transfer: the lagging behavior, Int. J. Thermophys, 36 (2015) 1416.

\bibitem{Shomali2021}	
Z. Shomali, R. Kovács, P. Ván, I.V. Kudinov, and J. Ghazanfarian, Recent Progresses and Future Directions of Lagging Heat Models in Thermodynamics and Bioheat Transfer, Continuum Mechanics and Thermodynamics, 34 (2022) 637.

\bibitem{Antaki2005}
P. J. Antaki, New Interpretation of Non-Fourier Heat Conduction in Processed Meat, J. Heat Transf, 127 (2005) 189.

\bibitem{Schelling2002}
P. K. Schelling, S. R. Phillpot, P. Keblinski, Phys. Rev. B, 65 (2002) 144306.

\bibitem{MXu2019}
M. Xu, Nonlocal heat conduction in suspended graphene, Phys. Lett. A. 383 (2019) 126017.

\bibitem{Abouelregal2025}
A. E. Abouelregal, M. Marin, and A. $\ddot{O}$chsner, A modified spatiotemporal nonlocal thermoelasticity theory with higher-order phase delays for a viscoelastic micropolar medium exposed to short-pulse laser excitation, Continuum Mechanics and Thermodynamics 37.1 (2025) 15.

\bibitem{RAGuyer1966}
R. A. Guyer, J.A. Krumhansl, Solution of the Linearized Phonon Boltzmann Equation, Phys. Rev. 148 (1966) 766.

\bibitem{RAGuyer1966-2}
R. A. Guyer, J.A. Krumhansl, Thermal conductivity, second sound, and phonon hydrodynamic phenomena in nonmetallic crystals, Phys. Rev., 148 (1966) 778.

\bibitem{Mongiovi2013}
M. S. Mongiov$\'{i}$ and M. Zingales, A non-local model of thermal energy transport: the fractional temperature equation. Int. J. Heat Mass Transf. , 67 (2013) 593.

\bibitem{Shomali2022}
M. H. Fotovvat and Z. Shomali, A time-fractional dual-phase-lag framework to investigate transistors with TMTC channels (TiS$_3$, In$_4$Se$_3$) and size-dependent properties, Micro and nanostructures, 168 (2022) 207304.

\bibitem{Salman2024}
Salman S.  Alsaeed, and Ahmed E. Abouelregal, Analysis of thermomechanical responses of functionally graded unbounded materials using an advanced dual‐phase delay heat transfer model with higher‐order fractional derivatives, ZAMM‐Journal of Applied Mathematics and Mechanics/Zeitschrift für Angewandte Mathematik und Mechanik (2024): e202400930.

\bibitem{Vermeersch2014}
B. Vermeersch, and A. Shakouri, Nonlocality in microscale heat conduction, arXiv preprint arXiv:1412.6555, 2014.

\bibitem{MXu2018}
M. Xu, A non-local constitutive model for nano-scale heat conduction, Int. J. Therm. Sci., 134 (2018) 594–600.

\bibitem{YJYu2016}
Y. J. Yu, C. L. Li, Z. N. Xue, X. G. Tian, The dilemma of hyperbolic heat conduction and its settlement by incorporating spatially nonlocal effect at nanoscale, Phys.
Lett. A. 380 (2016) 255.

\bibitem{DYTzou2010}
D. Y. Tzou, and Z. Y. Guo, Nonlocal behavior in thermal lagging, Int. J. Therm. Sci., 49(7) (2010) 1133.

\bibitem{DYTzou2011}
D.Y. Tzou, Nonlocal behavior in phonon transport, Int. J. Heat Mass Transf. 54 (2011) 475.

\bibitem{WPeng2024}
W. Peng, and B. Pan, Nonlocal dual-phase-lag thermoelastic damping analysis in functionally graded sandwich microbeam resonators utilizing the modified coupled stress theory, Mech. Based Des. Struct. Mach, 52(10) (2024) 7471.

\bibitem{LHai2024}
L. Hai and J. K. Dong, Nonlocal dual-phase-lag thermoelastic damping in small-sized circular cross-sectional ring resonators, Mech. Adv. Mater. Struct., 31.25 (2024) 7498.

\bibitem{Chawla2024}
A. Chawla, K. Sangeeta, and Sh. Surbhi, Wave Propagation in Nonlocal Dual-Phase-Lag Anisotropic Thermo-Elasticity, International Conference on Advances in Modern Age Technologies for Health and Engineering Science (AMATHE), IEEE (2024).

\bibitem{WYang2020}
W. Yang and Z. Chen, Nonlocal dual-phase-lag heat conduction and the associated nonlocal thermal-viscoelastic analysis, Int. J. Heat Mass Transf. 156 (2020) 119752.

\bibitem{Roya2023}
R. Baratifarimani, Z. Shomali, Implementation of nonlocal non-Fourier heat transfer for semiconductor nanostructures. Case Studies in Thermal Engineering, 54 (2024) 104015.

\bibitem{Subrina2009}
S. Subrina, D. Kotchetkov, and A. A. Balandin, Heat removal in silicon-on-insulator integrated circuits with graphene lateral heat spreaders. IEEE Electron device letters, 30(12) (2009) 1281.

\bibitem{Shomali2012}
J. Ghazanfarian and Z. Shomali, Investigation of dual-phase-lag heat conduction model in a nanoscale metal-oxide-semiconductor field-effect transistor, Int. J. Heat Mass Transf., 55(21-22) (2012) 6231.

\bibitem{Shomali20152}
Z. Shomali, J. Ghazanfarian, A. Abbassi, Investigation of bulk/film temperature-dependent properties for highly non-linear DPL model in a nanoscale device: the case with high-k metal gate MOSFET, Superlattices Microstruct., 83 (2015) 699.

\bibitem{Shomali2016}
Z. Shomali, A. Abbassi, J. Ghazanfarian, Development of non-Fourier thermal attitude for three-dimensional and graphene-based MOS devices, Appl. Therm. Eng., 104 (2016) 616.

\bibitem{Shomali2017}
Z. Shomali, B. Pedar, J. Ghazanfarian, A. Abbassi, Monte-Carlo Parallel Simulation of Phonon Transport for 3D Nano-Devices, Int. J. Therm. Sci., 114 (2017) 139.

\bibitem{2Shomali2017}
Z. Shomali, J. Ghazanfarian, A. Abbassi, 3-D Atomistic Investigation of Silicon MOSFETs, In Proceedings of CHT-17 ICHMT International Symposium on Advances in Computational Heat Transfer, ICHMT Digital Library Online, Begel House Inc., 2017.

\bibitem{Shomali2023}
Z. Shomali, An investigation into the reliability of newly proposed MoSi$_2$N$_4$/WSi$_2$N$_4$ field effect transistor: A monte carlo study, Micro and nanostructures, 182(3) (2023) 207648.

\bibitem{Yang2005}
R. Yang, G. Chen, M. Laroche, Y. Taur, Simulation of nanoscale multidimensional transient heat conduction problems using ballistic-diffusive equations and phonon Boltzmann equation, J. Heat Transfer, 127(3) (2005) 298.

\bibitem{Ghazanfarian2012}
J. Ghazanfarian, A. Abbassi,
Investigation of 2D transient heat transfer under the effect of dual-phase-lag model in a nanoscale geometry, International Journal of Thermophysics, 33 (2012) 552.

\bibitem{Ghazanfarian2009}
J. Ghazanfarian, A. Abbassi, \newblock{Effect of boundary phonon scattering on Dual-Phase-Lag model to simulate micro-and nanoscale heat conduction}, \newblock{Int. J. Heat Mass Transfer}, 52(15-16) (2009) 3706.

\bibitem{Basirat2006}
H. Basirat, J. Ghazanfarian, P. Forooghi, Implementation of dual-phase- lag model at different Knudsen numbers within slab heat transfer, In: Proceedings of International Conference on Modeling and Simulation (MS06), August, 2006, Konia, Turkey, 895899.

\bibitem{Zhmakin2023}
A. I. Zhmakin, Non-Fourier Heat Conduction: From Phase-Lag Models to Relativistic and Quantum Transport, Springer Nature, 2023.

\bibitem{Asheghi1997}
M. Asheghi, Y. K. Leung, S. S. Wong, K. E. Goodson, Phonon-boundary scattering in thin silicon layers, Applied Physics Letters, 71 (13) (1997) 1798.

\bibitem{Jiaung2008}
W. S. Jiaung, J. R. Ho, Lattice-Boltzmann modeling of phonon hydrodynamics, Physical Review E, 77(6) (2008) 066710.

\bibitem{Ghazanfarian2012-2}
J. Ghazanfarian, and A. Abbassi, Investigation of 2D transient heat transfer under the effect of dual-phase-lag model in a nanoscale geometry, International Journal of Thermophysics, 33 (2012) 552.

\bibitem{Dai2004}
W. Dai, L. Shen, R. Nassar, and T. Zhu, A stable and convergent three-level finite difference scheme for solving a dual-phase-lagging heat transport equation in spherical coordinates, Int. J. Heat Mass Transfer, 47 (2004)1817.

\end{thebibliography}
\end{document}